\documentclass{article}
\usepackage{amsfonts}

\usepackage{graphicx}
\usepackage{amsmath}

\input{tcilatex}

\begin{document}

\title{Hamiltonian and gradient properties of certain type of dynamical systems}
\author{A.K. Prykarpatsky\thanks{%
Institute of Mathematics, University of Mining and Metallurgy, Cracow,
Polland, Institute of Applied problem of Mechanics and Mathematics, National
Academy of Sciences{}} \ and V. V. Gafiychuk\thanks{
Mailling~address:~Institute of applied problems of Mechanics and
Mathematics, Nattional Academy of Scences,
Naukova~st,~3b,~~Lviv~79601,~Ukraine}}
\date{\today}
\maketitle

\begin{abstract}
From the sandpoint of neural network dynamics we consider dynamical system
of special type pesesses gradient (symmetric) and Hamiltonian
(antisymmetric) flows. The conditions when Hamiltonian flow properties are
dominant in the system are considered. A simple Hamiltonian has been studied
for establishing oscillatory patern conditions in system under consideration.

\noindent \textbf{keywords}: Neural network, Hamiltonian dynamical systems,
gradient dynamical system.
\end{abstract}

\section{Introduction and setting the problem.}

It is well known \cite{1} that synaptic connections in biological neural
networks are seldom symmetric since the signal sent by neurons along their
axons are sharp spikes and the relevant information is not contained in the
spikes themselves but in the so called firing rates, which depends on the
magnitude of the membrane potentials which governs all the process. On the
other hand it should be pointed out that the recent neurophysiological
observation of extremely low firing rates \cite{2} without some doubt on the
general usefulness of this notion as really the relevant neural variable.
Thereby one can use some natural continuous variables to describe neural
networks as dynamical systems of special structure like gradient (symmetric)
and Hamiltonian skew-symmetrical flows. This gives rise to making use of a
lot of methods and techniques for studying the structural stability of the
networks and the existence of so called coherent temporal structures fitting
for learning process.

Based on the considerations above one can introduce a class of nonlinear
dynamical systems

\begin{equation}
du/dt=K(u)  \label{1}
\end{equation}
where $M\ni u$ is some smooth finite-dimensional metrizible manifold $%
K:M\rightarrow T(M)$ is a vector field on $M$, modeling the information
transfer process in a biological neural network under regard. The question
is what conditions should be involved on the flow (\ref{1}) for it to be
represented as follows:

\begin{equation}
K(u)=-\limfunc{grad}V(u)-\vartheta (u)\nabla H(u),  \label{2}
\end{equation}
that is as a mixed sum of a gradient flow and a Hamiltonian flow on $M$.
Here $V:M\rightarrow \QTR{sl}{R}$ is the potential function and $%
H:M\rightarrow \QTR{sl}{R}$ is the Hamiltonian function relevant to the flow
(\ref{1}), $\limfunc{grad}:=g^{-1}(u)\nabla $, $\nabla :=\left\{ \frac{%
\partial }{\partial u}:u\in M\right\} $, $g:T(M)\times T(M)\rightarrow 
\QTR{sl}{R}_{+}$ is a Riemannian metrics and $\vartheta :$ $T^{\ast
}(M)\rightarrow T(M)$ is a Poisson structure on $M$.

Thus we need to find the corresponding metrics and Poisson structure on $M$
subject to which the representation (\ref{2}) holds on $M$. We shell dwell
on this topics in the proceeding chapter.

\section{Poissonian structure analysis}

Assume first that the representation (\ref{2}) holds, that is

\begin{equation}
-\vartheta (u)\nabla H(u)=K(u)+\limfunc{grad}V(u):=K_{V}(u)  \label{3}
\end{equation}
for all $u\in M$ and some $\vartheta $ and $g$ structures on $M$. This means
therefore that the constructed vector field (\ref{3}) is exactly
Hamiltonian. Thereby one has (\cite{3}) the expression

\begin{equation}
\vartheta ^{-1}(u)=\varphi ^{\prime }(u)-\varphi ^{\prime \ast }(u),
\label{4}
\end{equation}
where $\varphi \in T^{\ast }(M)$ is some nonsymmetric solution to the linear
determining equation

\begin{equation}
d\varphi /dt+K_{V}^{\prime \ast }\varphi =\nabla \mathcal{L}.  \label{5}
\end{equation}
Here, by definition, the flow $K_{V}$ is defined as

\begin{equation}
du/d\tau =K_{V}(u)  \label{6}
\end{equation}
and $\mathcal{L}:M\rightarrow \QTR{sl}{R}$ is a suitable smooth function
chosen for convenience when solving (\ref{4}). It is clear (see \cite{3})
that the symplectic structure (\ref{4}) doesn't depend on the choice of the
function $\mathcal{L}:M\rightarrow \QTR{sl}{R.}$

As the second step, assume that the metrics and Poisson structures on $M$
are given a priori. Then due to \ (\ref{5}) the following equation for
determining the potential function $V:M\rightarrow \QTR{sl}{R}$ holds:

\begin{equation}
\varphi ^{\prime }\cdot K+\varphi ^{\prime }\cdot \limfunc{grad}V+K^{\prime
\ast }\cdot \varphi +(\limfunc{grad}V)^{\prime \ast }\cdot \varphi =\nabla 
\mathcal{L},  \label{7}
\end{equation}
where the element $\varphi \in T^{\ast }(M)$ has been assumed also to be
known a priori as a solution to the equation (\ref{4}). The expression (\ref
{7}) is a linear second order equation in partial derivatives on the
potential function $V:M\rightarrow \QTR{sl}{R.}$ If this equation is
compatible, then its solution exists and the decomposition (\ref{2}) holds.

As one can check, the equation (\ref{7}) almost everywhere possesses a
solution for the vector $\psi =\limfunc{grad}V,$ that is the following
expression 
\begin{equation}
\limfunc{grad}V=\psi =g^{-1}\nabla V
\end{equation}
holds on $M$ for some $\psi \in T(M).$ Thereby, one gets

\begin{equation}
\nabla V=g\psi .  \label{9}
\end{equation}
Making use now of the well-known Volterra condition (see \cite{3}), $(\nabla
V)^{\prime \ast }\equiv (\nabla V)^{\prime }$, we obtain the following
criterion on the metrics $g:T(M)\times T(M)\rightarrow \QTR{sl}{R}_{+}$ :

\begin{equation}
(g\psi )^{\prime \ast }=(g\psi )^{\prime }.  \label{10}
\end{equation}
Since from (\ref{9}) also one follows that

\begin{equation}
\langle g\psi ,u_{x}\rangle =\langle \nabla V,u_{x}\rangle =dV/dx,
\label{11}
\end{equation}
the condition (\ref{11}) is evidently equivalent to such one:

\begin{equation}
(g\psi )^{\prime \ast }u_{x}-\frac{d}{dx}(g\psi )=0.  \label{12}
\end{equation}
Calculating the left handside expression of (\ref{12}) one gets the
following final result:

\begin{equation}
(g^{\prime \ast }u_{x}-g^{\prime }u_{x})\psi =g\psi ^{\prime }u_{x}-\psi
^{\prime }(gu_{x}),  \label{13}
\end{equation}
which is feasible at check, if the metrics is given. Otherwise, if this is
not the case, the linear expression (\ref{13}) determines a suitable metrics
as its solution subject to the mapping $g:T(M)\times T(M)\rightarrow 
\QTR{sl}{R}_{+}$ . As soon as the equation (\ref{13}) is compatible, its
solution exists defining a suitable metrics on the manifold $M$.

The results delivered above can be successfully applied to many interesting
dynamical systems modeling information processes in neural networks,
mentioned in introduction. Below we shall demonstrate some of them having
applications at studying coherent temporal structures.

\section{Coherent temporal structures formation.}

One considers a network with two groups of neuron $\left\{ x_{i}\in \QTR{sl}{%
R:\,}i\QTR{sl}{=}\overline{1,n}\right\} $ and \{$y_{j}\in \QTR{sl}{R}$%
\textsl{: }$j\QTR{sl}{=}\overline{1,m}$\}$,$ connected in such a way, that
inside both groups the synaptic strengths are symmetric, whereas between
groups they are antisymmetric. That is, neurons $\left\{ x\right\} $ are
excitatory to $\left\{ y\right\} $ and neurons $\left\{ y\right\} $ are
inhibitory to $\left\{ x\right\} .$ This model is expressed in the form (\ref
{2}), where 
\begin{eqnarray}
V &=&\overset{n}{\underset{i=1}{\sum }}\left( -\frac{1}{2}\beta
_{1}x_{i}^{2}+\beta _{2}\frac{x_{i}^{4}}{4}\right) +\frac{1}{2}\overset{n}{%
\underset{i,j=1}{\sum }}\beta _{i,j}^{(1)}x_{i}x_{j}  \label{14} \\
&&+\overset{m}{\underset{j=1}{\sum }}\left( -\frac{1}{2}\beta
_{4}y_{j}^{2}+\beta _{5}\frac{y_{j}^{4}}{4}\right) +\frac{1}{2}\overset{m}{%
\underset{i,j=1}{\sum }}\beta _{i,j}^{(2)}y_{i}y_{j},  \notag
\end{eqnarray}
\begin{equation}
H=\frac{1}{2}\left( \overset{n}{\underset{i=1}{\sum }}x_{i}^{2}+\overset{m}{%
\underset{j=1}{\sum }}y_{j}^{2}+\overset{n}{\underset{i=1}{\sum }}\overset{m%
}{\underset{j=1}{\sum }}w_{ij}x_{i}y_{j}\right)  \label{14a}
\end{equation}
with the standard metrics $g=\mathbf{1,}$a skew-symmetric Poisson structure $%
\vartheta =J\in \limfunc{Sp}(\QTR{sl}{R}^{(n+m)}),$ $u=\left\{ (x,y)\in 
\QTR{sl}{R}^{\QTR{sl}{n}}\times \QTR{sl}{R}^{\QTR{sl}{m}}\right\} ,$%
\begin{equation}
g=\left( 
\begin{array}{ccc}
1 &  & 0 \\ 
& \ddots &  \\ 
0 &  & 1
\end{array}
\right) ,\text{ }J=\left( 
\begin{array}{cc}
0 & I_{(n,m)} \\ 
-I_{(n,m)} & 0
\end{array}
\right)  \label{15}
\end{equation}
or 
\begin{equation}
J=\left( 
\begin{array}{cc}
J_{(n)} & 0 \\ 
0 & J_{(m)}
\end{array}
\right)  \label{15a}
\end{equation}
with constant $\beta $ and elements $w_{ij}$ being parameters, $%
I_{(n,m)}=\{\delta _{ij}:i=\overline{1,n},\;j=\overline{1,m}\},$ $J_{(n)}$
and $J_{(m)}$ \ being some skew-symmetric matrices.

It is worth to mention here that the representation (\ref{2}) with
structures (\ref{15}) is not unique and some other solutions to the equation
(\ref{13}) can be found.

This system (\ref{14}) as we shall demonstrate below possesses a so called
coherent-temporal structure important for studying learning processes in
biological neural networks.

Assume for simplicity that all $\beta $ - parameters are proportional to a
small enough parameter $\varepsilon >0$, that is $\left\{ \beta \right\}
\simeq \left\{ \varepsilon \beta \right\} $ and consider first our flow (\ref
{2}) at $\varepsilon =0$. It is easy to see that our model then possesses a
closed orbit in the space of $\left\{ x\right\} $ and $\left\{ y\right\} $ -
parameters, say $\sigma :\QTR{sl}{S}^{\QTR{sl}{1}}\rightarrow M=\QTR{sl}{R}^{%
\QTR{sl}{n}}\times \QTR{sl}{R}^{\QTR{sl}{m}},$ satisfying the equation 
\begin{equation}
d\sigma /d\tau =-J\nabla H(\sigma )  \label{16}
\end{equation}
for all $\tau \in \QTR{sl}{S}^{\QTR{sl}{1}}$. Moreover, the Hamiltonian
function $H:M\rightarrow \QTR{sl}{R}$ in (\ref{14a}) is a conservation law
of (\ref{16}). Take now $\varepsilon \neq 0$; then one can state ( \cite{4})
that there exists a function $H_{\varepsilon }:M\rightarrow \QTR{sl}{R}$,
such, that for some closed orbit $\sigma _{\varepsilon }:\QTR{sl}{S}^{%
\QTR{sl}{1}}\rightarrow M$ this function $H_{\varepsilon }:M\rightarrow 
\QTR{sl}{R}$ be a constant of motion (not a conservative quantity), that is
for all small enough $\varepsilon >0$%
\begin{equation}
dH_{\varepsilon }(\sigma _{\varepsilon })/dt=O(\varepsilon ^{2})  \label{17}
\end{equation}
as $\varepsilon \rightarrow 0$. Then one can formulate the following
proposition about the existence of a limiting cycle in our model at $%
\varepsilon >0$ small enough.

\textbf{Proposition.} Let our model possess at small enough $\varepsilon >0$
a smooth constant of motion $H_{\varepsilon }:M\rightarrow \QTR{sl}{R}$ and
a closed $\varepsilon $-deformed orbit $\sigma _{\varepsilon }:\QTR{sl}{S}^{%
\QTR{sl}{1}}\rightarrow $ $\QTR{sl}{R}$. Moreover, at $\varepsilon =0$ the
constant of motion $H_{0}:M\rightarrow \QTR{sl}{R}$ is a first integral of
the model in the neighborhood of the orbit $\sigma _{0}$. Then a necessary
condition for the existence of a limiting cycle at $\varepsilon >0$ small
enough is vanishing the following circular integral: 
\begin{equation}
\underset{\QTR{sl}{S}^{\QTR{sl}{1}}}{\oint }\left\langle \nabla H_{0}(\sigma
_{0}),\limfunc{grad}V(\sigma _{0})\right\rangle dt=0.  \label{18}
\end{equation}

Having substituted expression (\ref{14}) into (\ref{18}), one finds
numerical constraints on the parameters locating our closed orbit $\sigma
_{0}:\QTR{sl}{S}^{\QTR{sl}{1}}\rightarrow M$ in the phase space $M$.
Thereby, we can localize this way possible coherent temporal patterns
available in our neuron network under study.

Using this approach let us consider the equation of motion on the variables $%
(x,y)\in \QTR{sl}{R}^{n+m}.$ The Lagrangian equation corresponding to
potential (\ref{14}), Hamiltonian (\ref{14a}) and matrix (\ref{15a}) can be
represented as 
\begin{equation}
\left( 
\begin{array}{c}
\overset{..}{\mathbf{x}} \\ 
\overset{..}{\mathbf{y}}
\end{array}
\right) +W\left( 
\begin{array}{c}
\mathbf{x} \\ 
\mathbf{y}
\end{array}
\right) =0,  \label{19}
\end{equation}
where colums $\mathbf{x}=\{x_{1,}...x_{n}\}^{T\text{ \ }}$, $\mathbf{y}%
=\{y_{1,...,}y_{m}\}^{T},$ and matrix $W=\left( 
\begin{array}{cc}
A_{1} & B_{1} \\ 
A_{2} & B_{2}
\end{array}
\right) $, \ $A_{1}=J_{n}^{2}+J_{n}wJ_{m}w^{T}$, $%
A_{2}=J_{m}^{2}+J_{m}w^{T}J_{n}w,$ $B_{1}=J_{n}^{2}w+J_{n}wJ_{m}$, $%
B_{2}=J_{m}^{2}w+J_{m}w^{T}J_{n}$, $w=\{w_{ik}\}$. A\ solution to the matrix
equation (\ref{19}) can be represented as $(\mathbf{x,y})^{T}=\mathbf{a}\exp
(i\lambda t),$ with $\lambda \in \mathbb{C}$ \ being nontrivial only if the
following determinant 
\begin{equation}
\left| -\lambda ^{2}g+W\right| =0  \label{20}
\end{equation}
is equal to zero.

Equation (\ref{20}) is one of the degree $\ n+m$ subject to $\lambda ^{2}\in 
\mathbb{C}$ and determines 2($n+m)$ eigen frequencies $\omega _{s}=\left\{
\pm \omega _{1},...,\pm \omega _{m+n}\right\} .$ In this case the solution
gets the form 
\begin{equation}
\left( 
\begin{array}{c}
\mathbf{x}_{s} \\ 
\mathbf{y}_{s}
\end{array}
\right) =\mathbf{a}_{s}\exp (i\omega _{s}t)\mathbf{+a}_{s}^{\ast }\exp
(-i\omega _{s}t).  \label{21}
\end{equation}
Amplitudes $\mathbf{a}_{s}=\{a_{s,1},...,a_{s,n+m}\}^{T}$ must satisfy the
matrix equation

\begin{equation}
\left( -\omega _{s}^{2}g+W\right) \mathbf{a}_{s}=0.  \label{22}
\end{equation}
If we take that amplitudes $a_{s1}=a_{s1}^{\ast }=1$ for any $s\in \overline{%
1,n+m}$ solve (\ref{22}), we can get coefficients $K_{si}\,$of the
distribution of amplitudes relative to any frequency $\omega _{s}$. For this
case the solution (\ref{21}) can be represented as

\begin{eqnarray}
x_{sj} &=&a_{s1}K_{s,j}\exp (i\omega _{s}t)\mathbf{+}a_{s,1}^{\ast
}K_{s,j}\exp (-i\omega _{s}t)  \notag \\
&=&K_{s,i}\exp (i\omega _{s}t)\mathbf{+}K_{s,i}\exp (-i\omega _{s}t),
\label{23}
\end{eqnarray}

\begin{equation}
y_{sj}=K_{s,j+n}\exp (i\omega _{s}t)\mathbf{+}K_{s,j+n}\exp (-i\omega _{s}t).
\label{23a}
\end{equation}
\bigskip Here $K_{s,i},K_{s,j+n}$ are given constants depending on the
frequencies $\omega _{s}.$ Consider now the scalar product 
\begin{eqnarray*}
\left\langle \nabla H_{0}(\sigma _{0}),\nabla V(\sigma _{0})\right\rangle 
&=&\left( x_{1}+\overset{m}{\underset{j=1}{\sum }}w_{1j}y_{j}\right) \left(
-\beta _{1}x_{1}+\beta _{2}x_{1}^{3}+\frac{1}{2}\overset{n}{\underset{j=1}{%
\sum }}\beta _{1,j}^{(1)}x_{j}\right) +... \\
&&+\left( x_{n}+\overset{m}{\underset{j=1}{\sum }}w_{nj}y_{j}\right) \left(
-\beta _{1}x_{n}+\beta _{2}x_{n}^{3}+\frac{1}{2}\overset{n}{\underset{j=1}{%
\sum }}\beta _{n,j}^{(1)}x_{j}\right) + \\
&&+\left( y_{1}+\overset{n}{\underset{i=1}{\sum }}w_{i1}x_{i}\right) \left(
-\beta _{1}y_{1}+\beta _{2}y_{1}^{3}+\frac{1}{2}\overset{m}{\underset{j=1}{%
\sum }}\beta _{1,j}^{(2)}y_{j}\right) +... \\
&&+\left( y_{n}+\overset{n}{\underset{j=1}{\sum }}w_{in}x_{i}\right) \left(
-\beta _{1}y_{m}+\beta _{2}y_{m}^{3}+\frac{1}{2}\overset{m}{\underset{j=1}{%
\sum }}\beta _{n,j}^{(2)}y_{j}\right) .
\end{eqnarray*}
Having substituted solution (\ref{23}) into last expression and integrated
it along the period $T=2\pi /\omega _{s}$ we get a hyperplane which
determines the parameters of our neural network model:

\begin{eqnarray*}
&&\left( x_{1}+\overset{m}{\underset{j=1}{\sum }}w_{1j}K_{s,n+j}\right)
\left( -\beta _{1}+\frac{3}{4}\beta _{2}+\frac{1}{2}\overset{n}{\underset{j=1%
}{\sum }}\beta _{1,j}^{(1)}K_{s,j}\right) +... \\
&&+\left( K_{s,n}+\overset{m}{\underset{j=1}{\sum }}w_{nj}K_{s,n+j}\right)
\left( -\beta _{1}K_{s,n}+\frac{3}{4}\beta _{2}K_{s,n}^{3}+\frac{1}{2}%
\overset{n}{\underset{j=1}{\sum }}\beta _{n,j}^{(1)}K_{s,j}\right) +...
\end{eqnarray*}
\begin{equation*}
+\left( K_{s,n+1}+\overset{n}{\underset{i=1}{\sum }}w_{i1}K_{s,i}\right)
\left( -\beta _{3}K_{s,1+n}+\frac{3}{4}\beta _{4}K_{s,1+n}^{3}+\frac{1}{2}%
\overset{m}{\underset{j=1}{\sum }}\beta _{1,j}^{(2)}K_{s,j+n}\right) +...
\end{equation*}
\begin{equation*}
+\left( K_{s,n+m}+\overset{n}{\underset{j=1}{\sum }}w_{in}K_{s,i}\right)
\left( -\beta _{3}K_{s,m+n}+\frac{3}{4}\beta _{4}K_{s,m+n}^{3}+\frac{1}{2}%
\overset{m}{\underset{j=1}{\sum }}\beta _{n,j}^{(2)}K_{s,j+n}\right) =0
\end{equation*}
Thus, as the index $s$ changes from $s=1$ to $n+m,$ we can get in the
general case $n+m$ dimensional submanifolds determining parameters \{$\beta
_{1}$, $\beta _{2}$, $\beta _{3}$, $\beta _{4}$, $\beta _{n,j}^{(1)}$, $%
\beta _{n,j}^{(2)}$\}, at which the chosen oscillatory structure will
persist for all $t\in \QTR{sl}{R}_{+},$ thereby realizing a stable neural
network and related with it the temporal patter under study.

\end{document}